\documentclass[journal]{IEEEtran}
\usepackage{amsmath,amsfonts}
\usepackage[x11names]{xcolor}
\usepackage{algorithmic}
\usepackage{array}
\usepackage[caption=false,font=normalsize,labelfont=sf,textfont=sf]{subfig}
\usepackage{textcomp}
\usepackage{stfloats}
\usepackage{url}
\usepackage{verbatim}
\usepackage{graphicx}
\usepackage{hyperref}
\hypersetup{
    colorlinks=false,
}
\def\BibTeX{{\rm B\kern-.05em{\sc i\kern-.025em b}\kern-.08em
    T\kern-.1667em\lower.7ex\hbox{E}\kern-.125emX}}
\usepackage{balance}
\usepackage{cite}

\begin{document}

\title{KP-A: A Unified Network Knowledge Plane for Catalyzing Agentic Network Intelligence}

\author{Yun Tang,~\IEEEmembership{Member,~IEEE}, Mengbang Zou, Zeinab Nezami,~\IEEEmembership{Member,~IEEE}, \\Syed Ali Raza Zaidi,~\IEEEmembership{Senior Member,~IEEE}, Weisi Guo,~\IEEEmembership{Senior Member,~IEEE}

\thanks{Yun Tang, Mengbang Zou, and Weisi Guo are with Cranfield University, UK. Emails: \{yun.tang, m.zou, weisi.guo\}@cranfield.ac.uk. 
Zeinab Nezami (z.nezami@leeds.ac.uk) and Syed Ali Raza Zaidi (s.a.zaidi@leeds.ac.uk) are with the University of Leeds, UK. 
This work is supported by EPSRC CHEDDAR: Communications Hub For Empowering Distributed ClouD Computing Applications And Research (EP/X040518/1) (EP/Y037421/1). Corresponding author: Yun Tang. AI tools have been used for clarity and grammar revision.
}
}

\markboth{Journal of \LaTeX\ Class Files,~Vol.~14, No.~8, August~2021}%
{Shell \MakeLowercase{\textit{et al.}}: A Sample Article Using IEEEtran.cls for IEEE Journals}


\maketitle

\begin{abstract}
The emergence of large language models (LLMs) and agentic systems is enabling autonomous 6G networks with advanced intelligence, including self-configuration, self-optimization, and self-healing. However, the current implementation of individual intelligence tasks necessitates isolated knowledge retrieval pipelines, resulting in redundant data flows and inconsistent interpretations. Inspired by the service model unification effort in Open-RAN (to support interoperability and vendor diversity), we propose KP-A: a unified Network \textbf{K}nowledge \textbf{P}lane specifically designed for \textbf{A}gentic network intelligence. By decoupling network knowledge acquisition and management from intelligence logic, KP-A streamlines development and reduces maintenance complexity for intelligence engineers. By offering an intuitive and consistent knowledge interface, KP-A also enhances interoperability for the network intelligence agents. We demonstrate KP-A in two representative intelligence tasks: live network knowledge Q\&A and edge AI service orchestration. All implementation artifacts have been open-sourced to support reproducibility and future standardization efforts.
\end{abstract}


\section{Introduction}
\IEEEPARstart{T}{he} sixth generation (6G) of mobile networks, as envisioned in the IMT-2030 framework, aims to revolutionize wireless communication by offering ultra-high data transmission rates, ultra-low latency, greater connection capacity, and ubiquitous connectivity to enable a wide array of novel applications and services, including immersive communications, integrated sensing and communication, and hyper-reliable low-latency communications~\cite{ITU-R-M.2160-0}. 
To meet the complex and dynamic service provisioning requirements of these use cases, 6G networks must incorporate advanced self-management capabilities such as self-configuration, self-optimization, and self-healing, which enable networks to autonomously adapt to changing conditions with minimal human intervention~\cite{ETSI-GS-ZSM-012}.

The integration of advanced Artificial Intelligence (AI), particularly large language models (LLMs) or LLM-powered agentic systems, into network operations is critical to realizing these autonomy capabilities. In recent years, researchers have widely adopted LLMs and LLM-powered agents in various components of the telecom network, such as configuration management, anomaly detection, network security, network function code generation, network simulation, user intent profiling, and service orchestration.

\begin{figure*}
    \centering
    \includegraphics[width=\textwidth]{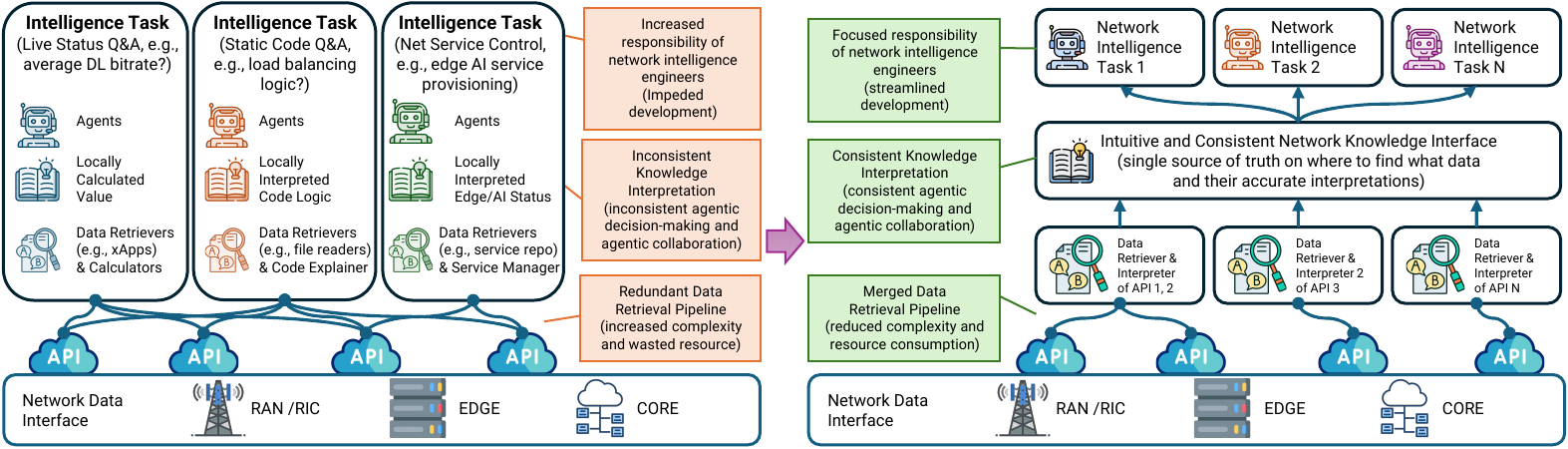}
    \caption{Comparison of knowledge engineering approaches (left: isolated and inconsistent; right: unified and consistent) for network intelligence tasks (and intelligence engineers) and the envisioned benefits of KP-A.}
    \label{fig:knowledge-engineering-approach-comparison}
\end{figure*}

\begin{figure*}
    \centering
    \includegraphics[width=\textwidth]{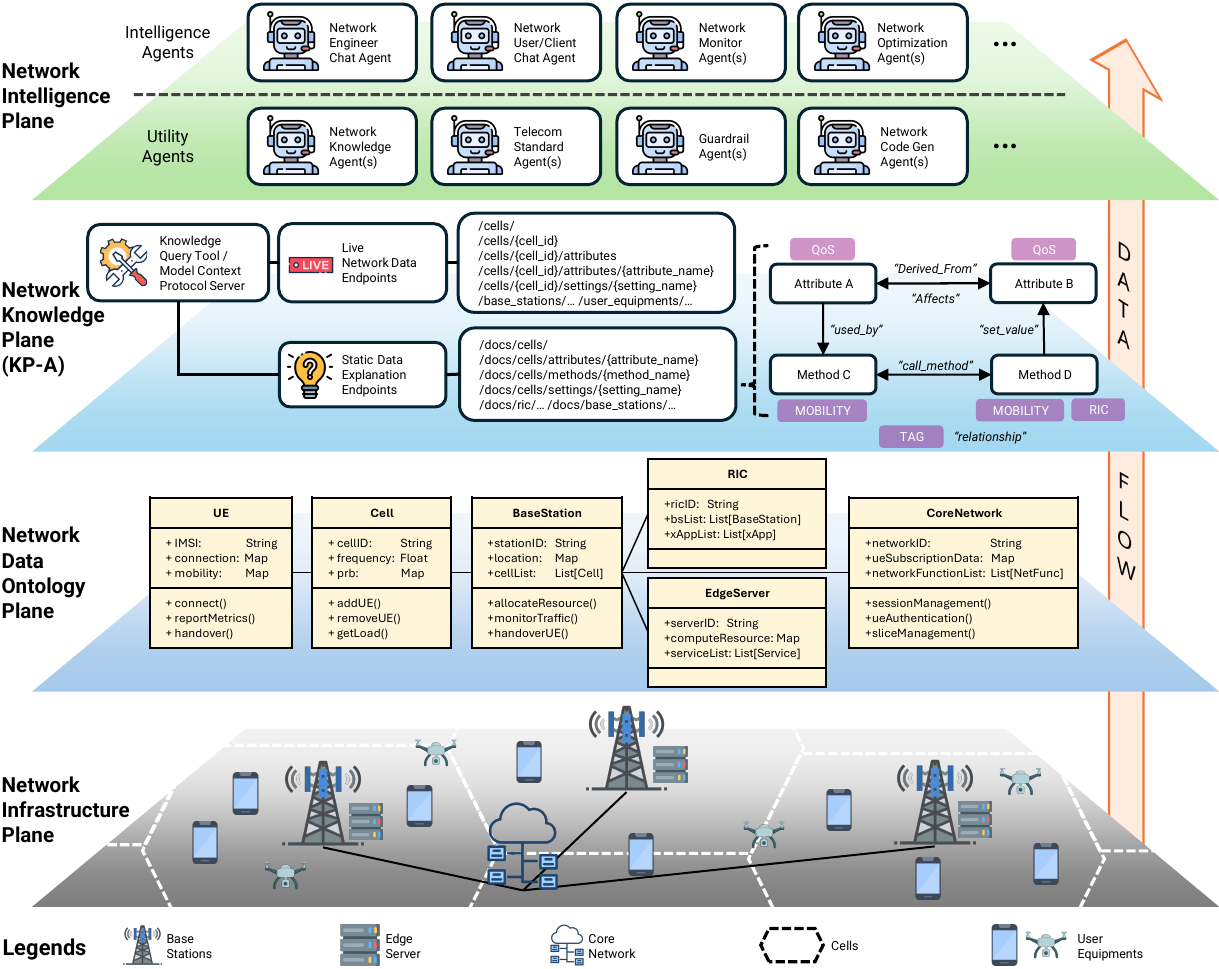}
    \caption{Overview of KP-A and its relative logical position in the autonomous 6G network. Note that in Software-Defined Networks, there can be multiple knowledge provisioning servers across the network software cluster, and the intelligence agents can run wherever applicable (detailed in Section~\ref{sec:network-knowledge-plane}).}
    \label{fig:knowledge-plane-architecture}
\end{figure*}

However, despite the sound benefits, integrating all the above proposed LLM pipelines and agentic frameworks into one telecom network from the network service provider's point of view is impractical. One of the primary issues is the lack of a unified knowledge plane, and as a result, network intelligence engineers need to implement additional knowledge retrieval and interpretation mechanisms. As illustrated in Fig.~\ref{fig:knowledge-engineering-approach-comparison}, such a fragmented practice leads to redundancy, inconsistent interpretations of network data, and increased development and maintenance overhead. For instance, there can be a scenario where ten LLM-powered intelligence pipelines deploy ten xApps independently but subscribe to the same event from the underlying network infrastructure, resulting in severe redundancy, suboptimal network performance, and, more importantly, inconsistent information interpretations.

To address these challenges, we propose a unified network \textbf{K}nowledge \textbf{P}lane for \textbf{A}gentic intelligence (KP-A) situated between the underlying Network infrastructure plane and the upper Intelligence Plane. This intermediary layer serves as a unified source of truth, ensuring consistent and intuitive information retrieval protocols and interpretation across intelligent applications. By unifying knowledge management, KP-A reduces redundancy, harmonizes interpretations, and streamlines the development of agentic intelligence frameworks.
Our contributions in this paper are summarized as follows:

\begin{itemize}
\item We propose KP-A, a unified network knowledge plane to facilitate consistent and intuitive knowledge exploration and retrieval by network intelligence agents.
\item We implement and demonstrate our design through two representative network intelligence tasks: (i) live network knowledge Q\&A and (ii) edge AI service provisioning.
\item We open-source all the source code (O-RAN network simulator with KP-A and intelligence agents) to foster further research and standardization in this field.
\end{itemize}

\section{Related Works}

\subsection{Knowledge Plane and Knowledge Defined Network}
The concept of \textit{Knowledge Plane} (KP) dates back to 2003 by David Clark~ \cite{clark2003knowledge}, which initiated the proliferation of \textit{Knowledge-defined Networks} (KDN) \cite{mestres2017knowledge}. The essence of KDN is to construct a global and unified perspective on the data and controls of the underlying physical/software-defined network to enable traditional AI/ML-based network automation. 

Entering the era of advanced network intelligence powered by LLMs and agents, we adopt the design philosophy of the knowledge plane and propose KP-A, which is specifically designed for agentic network intelligence.

\subsection{Knowledge Demands for LLM/Agent-based Intelligence}

Novel applications in various network management processes have been designed, demanding knowledge from different sources within the network infrastructure:

\subsubsection{LLMs as Network Data/Event Explainers}
Mekrache et al.~\cite{mekrache2024combining} propose an anomaly explanation pipeline that combines XGBoost detection, SHAP-based feature attribution, and the LLaMA2 large language model to generate human-friendly recommendations by leveraging external operational knowledge via in-context learning.
Similarly, Long et al.~\cite{long20246g} develop an LLM-driven network health reporting system that integrates multimodal logs, performance metrics, and external knowledge such as fault trees and remedy histories to enable accurate fault diagnosis and resolution.

\subsubsection{LLMs as User Intent Translators}
Lira et al.~\cite{lira2024large} introduce LLM-NetCFG, a system that translates natural language intents into network device configurations, thereby reducing manual management effort through model-based knowledge reasoning.
Similarly, Tong et al.~\cite{tong2025wirelessagent} present WirelessAgent, a multi-agent framework that leverages retrieval-augmented generation with 3GPP specifications and real-time network status to interpret user intents and autonomously manage network slicing.

\subsubsection{LLMs as Network Code Generators}
Jiang et al.~\cite{jiang2024large} propose CommLLM, which integrates multi-agent retrieval, collaborative planning, and reflective evaluation over domain-specific knowledge bases—including communication standards, technical documents, and research papers—to generate source code for semantic communication models.

\subsubsection{LLMs as Network Simulation Drivers}
Rezazadeh et al.~\cite{rezazadeh2025toward} introduce a multi-agent ns-3 simulation framework in which LLM agents leverage vectorized network documentation and parameters to automate the generation, testing, and interpretation of simulation scenarios using the ns-3 simulator.

For all the application patterns above, having access to external knowledge sources is essential to ensure the trustworthiness and effectiveness of LLMs and agent-based systems, particularly due to the following reasons: (1) new information and knowledge are continuously generated, making static models quickly outdated; (2) LLMs are inherently limited by their training data, and frequent retraining or fine-tuning to incorporate new knowledge is substantially more expensive than in-context learning; and (3) the internal knowledge retrieval process within LLMs, driven by the attention mechanism, is inherently probabilistic; meanwhile external, verifiable knowledge sources can serve as critical ground truth references that guide models towards accurate and reliable responses.

\subsection{Knowledge Engineering for LLMs and Agentic Systems}
There are three major engineering practices to organize external knowledge based on the nature of the knowledge:

\subsubsection{Semantic Databases for Static Factual Knowledge} 

\begin{figure*}
    \centering
    \includegraphics[width=\textwidth]{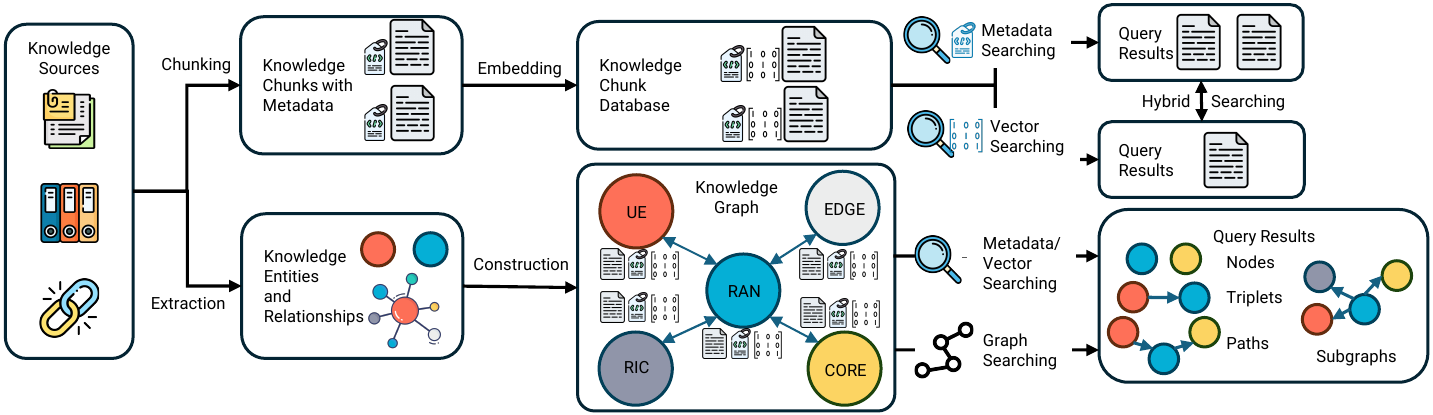}
    \caption{Common knowledge engineering approaches for RAG: (top) knowledge chunking and semantic matching, and (bottom) knowledge graph construction and graph search.}
    \label{fig:common_knowledge_engineering_approaches}
\end{figure*}

Semantic databases organize factual knowledge by chunking documents into smaller units, embedding them into semantic vectors (e.g., via OpenAI’s \textit{text-embedding-3-small}), and storing them with metadata for efficient retrieval. As shown in the top path of Fig.~\ref{fig:common_knowledge_engineering_approaches}, retrieval combines metadata filtering with semantic matching based on vector similarity, enabling scalable and precise retrieval-augmented generation (RAG) pipelines.

\subsubsection{Knowledge Graphs for Static Structural Knowledge}


Knowledge graphs preserve ontological relationships among entities that would otherwise be lost in naive text chunking (bottom path of Fig.~\ref{fig:common_knowledge_engineering_approaches}). For instance,~\cite{xu2024retrieval} constructs a dual-level KG from support tickets to improve QA accuracy, while GRAG~\cite{hu2024grag} retrieves subgraphs via ego-graph-based strategies and integrates both text and graph views for enhanced multi-hop reasoning. In industrial contexts, KG-enhanced RAG frameworks~\cite{bahr2025knowledge} formalize FMEA data into multi-relational graphs, supporting symbolic and numerical reasoning over failure-cause-mitigation chains via efficient graph traversals.

\subsubsection{Digital Twins for Dynamic Operational Knowledge}

While the above offer robust solutions for encoding and retrieving static knowledge, whether factual or structural, they fall short in scenarios requiring real-time insights from continuously evolving environments. On the other hand, digital twins have been widely adopted in industrial settings as a paradigm for integrating dynamic operational knowledge, where contextual updates, sensor data, and inference mechanisms co-evolve with the physical systems they represent.

For instance, Wang et al.~\cite{wang2025radio} introduce the Radio Environment Knowledge Pool (REKP), designed to support 6G digital twin channels by capturing how wireless signals behave in complex, real-world environments. REKP works like a dynamic memory system that learns and stores the relationship between physical surroundings (like buildings or obstacles) and signal behaviors (such as fading or path loss), using data from sensors and simulation tools. This allows the system to predict and adapt to changes in signal quality with greater accuracy and speed. One use case simulates a street-level environment, showing how REKP updates its understanding of signal loss as a receiver moves through areas with different building densities and obstructions.

\subsection{Current Challenges}

Despite many successful network intelligence applications with LLMs and agentic systems, from the literature review, we identify several challenges in current knowledge engineering practices. 

\subsubsection{Redundant Knowledge Pipeline}

Agentic intelligence tasks typically implement their logic for retrieving and interpreting network data individually. This fragmented approach can lead to 1) redundant subscriptions where there can be numerous xApps subscribing to the same network events, leading to unnecessary overhead in both the control and data planes; and 2) hindered development due to increased engineering effort to develop and maintain parallel, uncoordinated knowledge pipelines across diverse tasks.

\subsubsection{Poor Reusability and Interoperability}
Existing knowledge artifacts are closely tied to their target applications and input data formats. This tight coupling thus inhibits reuse and generalization across different network intelligence tasks. For example, extensive prompt engineering and fine-tuning may be required to adapt to different knowledge interfaces designed for existing intelligence tasks. Orchestrating end-to-end workflows that involve multiple LLMs or agents can also become error-prone and operationally brittle as the agents may be forced to communicate in different knowledge ``accents'' and query knowledge with different protocols. In addition, proprietary knowledge formats or APIs also hinder interoperability across vendors and operators, leading to vendor lock-in.

\subsubsection{Inconsistent Interpretations}

The absence of a unified source of truth can lead to diverging decisions or actions triggered by the same underlying network event, due to isolated semantic interpretations or retrieval mechanisms of individual intelligence tasks.

\section{KP-A: Network Knowledge Plane for Agents}\label{sec:network-knowledge-plane}

\subsection{Design Requirements}

To ensure the KP-A meets the practical needs of real-world 6G telecom deployments, we outline the following \textbf{ten} key design requirements.

\subsubsection{Schema}
The KP-A should provide an industry-standard-aligned schema or ontology to represent dynamic network knowledge, such as state, topology, and categorizations. This ensures semantic consistency across networks, vendors, and intelligence pipelines. The schema should be extensible to accommodate evolving 6G concepts.

\subsubsection{Freshness}
The KP-A should provision real-time or near-real-time network states through continuous streaming or event-driven mechanisms. Temporal versioning should be supported to enable time-based analysis, historical debugging, or replay of network events.

\subsubsection{Coverage}
The KP-A should provision the broadest possible range of network knowledge required by the Intelligence Plane. This includes granular, real-time operational data (e.g., current Channel Quality Indicator (CQI), Cell Individual Offset (CIO), and PRB allocations per UE), high-level orchestration states (e.g., execution logs and policies of xApps and service orchestrators), and even static or procedural artifacts (e.g., source code for scheduling, handover, and load-balancing algorithms). By making such comprehensive and cross-domain knowledge uniformly accessible, the KP-A can support both low-level data retrieval and high-level reasoning, promoting deeper situational awareness and advanced network autonomy.

\subsubsection{Semantic Enrichment}
The KP-A should provision lightweight insights wherever possible to bootstrap high-level reasoning from low-level metrics (e.g., congestion prediction, anomaly detection) with rule engines or AI/ML pipelines.

\subsubsection{Maintainability}
The KP-A should decouple knowledge acquisition from its consumption, enabling multiple intelligent agents to reuse shared network insights. It should expose modular and intuitive APIs for seamless integration with upstream data sources and downstream intelligence consumers.

\subsubsection{Security}
The KP-A should support multi-tenant architectures with fine-grained, role-based access control. This ensures secure and isolated access to knowledge subsets while enforcing privacy and compliance policies such as data masking or anonymization.

\subsubsection{Performance}
The KP-A should scale horizontally to accommodate the massive volume and velocity of knowledge generated by dense deployments of heterogeneous 6G elements, such as user equipment, small cells, IoT devices, and edge nodes. Scalability in query handling is essential to ensure that knowledge flows remain timely and actionable, especially for real-time decision-making.

\subsubsection{Auditability}
The KP-A should log all knowledge updates and accesses to support traceability, compliance audits, and root-cause analysis. Additionally, it should enable consuming agents to expose which knowledge artifacts informed their decisions, supporting explainability.

\subsubsection{Resilience}
The KP-A should operate reliably in the presence of network failures, latency spikes, or cyber threats. Redundancy and failover mechanisms should be incorporated to ensure high availability and fault tolerance in mission-critical deployments.

\subsubsection{Observability}:
The KP-A should include toolings for developers and operators, such as monitoring dashboards and query testing utilities. Observability mechanisms should report metrics related to data freshness, system health, and query performance.

\subsection{KP-A Design Overview}

Fig.~\ref{fig:knowledge-plane-architecture} illustrates the architectural design of the unified Network Knowledge Plane for Agentic intelligence (KP-A) and its integration within the autonomous 6G network. The design adopts a layered approach, with the KP-A positioned between the lower-level Network Data Ontology Plane and the upper-level Network Intelligence Plane. Its primary roles are to abstract and organize dynamic network information and provision intuitive, consistent, and interpretable textual knowledge, serving as a unified and reliable foundation for developing agentic network intelligence.

At the bottom of the stack lies the \emph{Network Infrastructure Plane}, composed of physical and virtual elements such as user equipment (UE), base stations, edge servers, and core network components. These elements form the operational substrate of the 6G network, continuously generating telemetry and control data.

Immediately above, the \emph{Network Data Ontology Plane} defines structured object-oriented models for these network entities. Each model, such as UE, Cell, Base Station, RIC, or Core Network, includes attributes (e.g., frequency, connection state, resource allocation) and logical methods (e.g., \texttt{connect()}, \texttt{getLoad()}, \texttt{handoverUE()}). This plane collects and organizes the raw data flows from the network infrastructure plane (e.g., via O-RAN or OAM interfaces) with a standardized ontology that enables intuitive, deterministic, and structured knowledge exploration and retrieval for the layers above.

The core of the architecture is the \emph{Network Knowledge Plane}—the KP-A itself. This plane transforms the fragmented data from the Data Ontology Plane into REST-style queryable knowledge constructs. It exposes two primary types of interfaces: \emph{live network data endpoints} and \emph{static data explanation endpoints}. The live endpoints provide real-time access to evolving network states, while the static endpoints document how the available attributes, methods, and configurations shall be interpreted by the agents. Beyond this, the KP-A introduces semantic structures, such as attribute and method relationships, to describe how various elements affect one another, facilitating deeper semantic understanding and iterative knowledge exploration. For example, a QoS attribute might be \texttt{derived\_from} another metric or \texttt{used\_by} a particular network method. These relationships form a lightweight knowledge graph that enables reasoning and context-aware querying by the intelligence agents.

Above the KP-A, the \emph{Network Intelligence Plane} hosts a diverse set of LLM-powered agents responsible for question-answering and decision-making (optimization, diagnostics, and network planning). These agents can be broadly categorized into \emph{intelligence agents}, which are tasked with high-level functions such as network engineering, client interaction, and optimization, and \emph{utility agents}, which focus on specific tasks like monitoring, standard enforcement, or guardrail enforcement. These agents access the KP-A through a dedicated knowledge query tool or model context protocol, allowing them to retrieve, interpret, and reason over dynamic network knowledge without needing to parse telemetry feeds or hardcode the configuration logic individually.

\section{Prototype Demonstration}

To demonstrate, we implemented and open-sourced a prototype knowledge plane based on our in-house lightweight RAN simulator~\cite{our_simulator}.

\subsection{Network Engineer Chat Agent}

\begin{figure}
    \centering
    \includegraphics[width=\columnwidth]{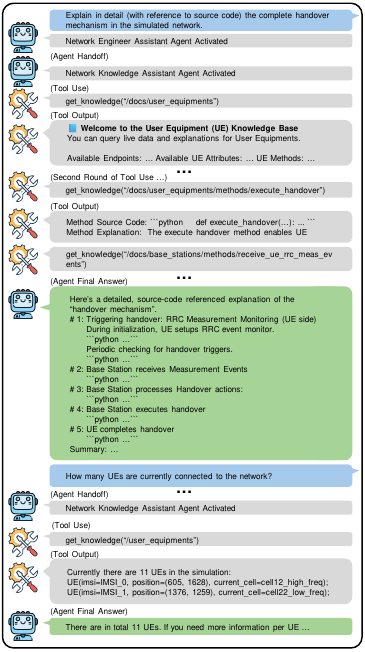}
    \caption{Agent querying the KP-A for static and dynamic network knowledge.}
    \label{fig:network-engineer-chat}
\end{figure}

Fig.~\ref{fig:network-engineer-chat} demonstrates how intelligence plane agents, such as the Network Engineer Chat Agent, seamlessly retrieve both static and dynamic knowledge from KP-A to answer complex queries. In this example, an engineer requests a detailed explanation of the handover mechanism in the simulated RAN environment—a query requiring source code-level insights across multiple simulated entities, including the UE, cell, base station, and RIC. Upon activation, the Network Engineer Chat Agent hands off the query to the Network Knowledge Agent, which is equipped with knowledge query tools to interact with KP-A. Through two rounds of knowledge exploration, the agent collects sufficient source code references and explanations to formulate a comprehensive response. The knowledge endpoints in KP-A are designed to be self-contained and interlinked. For example, the \texttt{/docs/ue/methods/execute\_handover} endpoint returns the actual code snippets (not pseudo-code), explanations on how the code shall be interpreted, and references to related attributes and methods. Queries for dynamic operational data, such as the number of UEs connected to the network, are resolved similarly through real-time knowledge endpoints.

This demonstration exemplifies how KP-A empowers agents with direct, structured, and consistent access to comprehensive network knowledge, thereby potentially facilitating explainability, root cause analysis, or rapid in-situ troubleshooting without requiring manual inspection of fragmented logs or raw data from the underlying data ontology plane and infrastructure plane.

\subsection{Edge AI Service Provisioning Agent}

\begin{figure}
    \centering
    \includegraphics[width=\columnwidth]{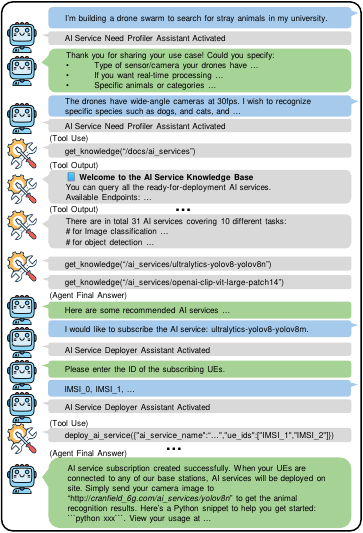}
    \caption{Agent querying KP-A for edge AI service provisioning.}
    \label{fig:ai-service-chat}
\end{figure}

Fig.~\ref{fig:ai-service-chat} illustrates the workflow of the Edge AI Service Provisioning Agent leveraging KP-A to facilitate AI service deployment for user equipment (UEs) at the network edge. In this example, a use case developer (network customer) initiates a conversation with the AI Service Need Profiler Assistant, outlining their objective of deploying a drone swarm equipped with wide-angle cameras to detect stray animals such as dogs and cats. The profiler agent collects deployment requirements, including sensor specifications, real-time processing needs, and target categories, before invoking the AI Service Knowledge Base to query the catalog of ready-to-deploy AI services. Upon retrieving the available services, it identifies and recommends suitable models such as YOLOv8 for real-time object detection. Following the developer’s service selection, the AI Service Deployer Assistant is activated to create AI service subscriptions for the specified UEs. The final response confirms successful subscription creation and provides operational details, including the AI inference endpoint URL and a Python code snippet to support immediate integration.

This demonstration exemplifies how KP-A extends beyond static knowledge retrieval to enable on-demand service provisioning and potentially general network control tasks.

\subsection{More Evaluation Examples}

More evaluation examples can be found in our project repository \cite{our_simulator}, containing 1) single attribute/method Q\&A conversations that require only one knowledge query for a single attribute or method of a network entity class (e.g., \textit{what's the current cell connected by the UE of ID: IMSI\_1?}); 2) multi-method logic Q\&A conversations that require multiple queries on different methods of a single network entity class (e.g., \textit{explain what happens when an UE powers up with code-level reference}); 3) cross-entity logic Q\&A conversations that require multiple queries on different methods of multiple network entity classes (e.g., \textit{explain the complete handover mechanism implementation in the simulated network with code-level reference}); and 4) knowledge endpoint reusability validation conversations which involves multiple rounds of Q\&A and the agent are expected to infer knowledge endpoints based on previous rounds.

\section{Conclusion}

In this paper, we introduced KP-A, a unified Network Knowledge Plane designed to overcome key challenges in developing agentic intelligence for 6G networks, including fragmented knowledge pipelines, redundant data flows, inconsistent interpretations, and hindered interoperability. By decoupling knowledge acquisition and management from intelligence logic, KP-A provides a reusable, consistent, and intuitive foundation for building diverse network intelligence applications. We demonstrated its practical value through two representative use cases: network engineer chat for source-code-referenced explanations and edge AI service provisioning for dynamic service orchestration.

Future work will focus on systematic evaluation, extending the schema coverage to encompass emerging 6G technologies such as integrated sensing and communication (ISAC), enabling cross-domain knowledge ingestion and orchestration, and potentially standardizing KP-A interfaces for widespread adoption in real-world telecom deployments. We believe KP-A will catalyze the next generation of trustworthy, explainable, and autonomous 6G networks by serving as a foundation for large-scale AI and agentic systems.



\bibliographystyle{IEEEtran}
\bibliography{references}

\begin{IEEEbiographynophoto}{YUN TANG (Member, IEEE)}
is a post-doctoral research fellow with Cranfield University from March 2024. He obtained his PhD in January 2023 at Nanyang Technological University, Singapore, and joined the University of Warwick as a research fellow from February 2023 to Mar 2024. His works have been published at several top-tier international conferences, such as IEEE ITSC, IEEE ICRA, and ASE, and in journals such as IEEE TSE and IEEE TIV. Currently, his research focuses on intelligence autonomy in future networks.
\end{IEEEbiographynophoto}

\begin{IEEEbiographynophoto}{MENGBANG ZOU}
is a post-doctoral research fellow with Cranfield University from 2024, founded by EPSRC 6G Future Communications Hubs in Distributed Computing. He obtained his PhD degree in February 2024 from Cranfield University, United Kingdom. His research interests are mainly in the dynamics of networked systems, complex networks, and machine learning.
\end{IEEEbiographynophoto}

\begin{IEEEbiographynophoto}{ZEINAB NEZAMI (Member, IEEE)}
Zeinab Nezami is a Postdoctoral Researcher at the University of Leeds, affiliated with the Schools of Computing and Electronic and Electrical Engineering. She contributes to CHEDDAR HUB, focusing on developing and standardizing GenAI agents for future communication systems and fostering collaboration among academia, industry, and global partners.
Her research interests include distributed intelligence, edge computing, and multi-agent systems. Zeinab also co-leads an AISI-funded project with Ericsson and NVIDIA on safety, governance, and cooperation among AI agents in telecom networks. She holds a PhD in Distributed Systems.

\end{IEEEbiographynophoto}

\begin{IEEEbiographynophoto}{SYED ALI RAZA ZAIDI (Senior Member, IEEE)}
is an Associate Professor (University Academic Fellow) at the University of Leeds, appointed under the prestigious 250 Great Minds Initiative. His research focuses on communication and networking for IoT, robotics, and autonomous systems. He has worked on high-impact projects, including a U.S. Army Research Lab-funded initiative on cognitive green wireless communication. He holds a PhD from the University of Leeds, funded by multiple competitive scholarships, and a B.Eng from NUST, where he earned the Rector’s Gold Medal. An active contributor to the academic community, he has served as a TPC member for major IEEE conferences and a reviewer for flagship journals. In addition to his academic work, he is a certified embedded systems and programming professional with certifications from Microsoft, IBM, and Sun Microsystems.
\end{IEEEbiographynophoto}

\begin{IEEEbiographynophoto}{WEISI GUO (Senior Member, IEEE)}
is a Professor of Human Machine Interface at Cranfield University and also with the Alan Turing Institute. Previously, he was with the University of Cambridge, the University of Sheffield, and the University of Warwick. He is the winner of several IEEE Best Paper and IET Innovation awards in communications and currently co-leads the EPSRC 6G Future Communications Hubs in Distributed Computing and the EPSRC Trustworthy Autonomous Systems Node in Security. He has published over 150 journal papers, mostly on networking and autonomy, with papers in Nature, Nature Communications, and Nature Machine Intelligence. He has served as editor to 3 IEEE and 1 Royal Society journals and contributes to emerging standards in wireless and AI. 
\end{IEEEbiographynophoto}

\vfill

\end{document}